 \documentclass[twocolumn]{webofc}

\usepackage[varg]{txfonts}   
\usepackage{hyperref}
\usepackage{url}
\hypersetup{colorlinks=true,citecolor=blue,urlcolor=blue,linkcolor=blue}
\usepackage{soul}

\newcommand{\PBF}{\ensuremath{\rm{PbF_2}}\xspace}
\newcommand{\PWO}{\ensuremath{\rm{PbWO_4}}\xspace}

\newcommand{\BGO}{BGO\xspace}
\newcommand{\tMCP}{\ensuremath{t_{\rm MCP}}\xspace}

\begin{document}
\title{Dual-readout calorimetry with homogeneous crystals}

\author{
  \firstname{Robert} \lastname{Hirosky}\inst{1}
    \fnsep\thanks{\email{hirosky@virginia.edu}} 
    \and
  \firstname{Thomas} \lastname{Anderson}\inst{1}
    \and
  \firstname{Grace} \lastname{Cummings}\inst{2}
    \and
  \firstname{Max} \lastname{Dubnowski}\inst{1}
    \and
    \firstname{Christian} \lastname{Guinto-Brody}\inst{1}
    \and
  \firstname{Yuxiang} \lastname{Guo}\inst{3}
    \and
  \firstname{Alexander} \lastname{Ledovskoy}\inst{1}
       \and
  \firstname{Daniel} \lastname{Levin}\inst{3}
    \and
  \firstname{Christopher} \lastname{Madrid}\inst{2}
    \and
  \firstname{Christopher} \lastname{Martin}\inst{1} 
     \and
  \firstname{Junjie} \lastname{Zhu}\inst{3}
}

\institute{The University of Virginia, Charlottesville VA
\and
           Fermi National Accelerator Lab, Batavia IL 
\and
           University of Michigan, Ann Arbor MI
          }

\abstract{

High resolution calorimetry with state-of-the-art energy resolution performance for both electromagnetic (EM) and hadronic signals can be achieved using the dual-readout (DR) technique, both in a homogeneous scintillating-crystal calorimeter and in a traditional fiber and absorber-based DR hadronic section.  We present results from the CalVision consortium studying the collection of Cerenkov and scintillation signals in PbWO$_4$ and BGO crystal samples exposed to 120\,GeV proton beams at the Fermilab Test Beam Facility, including proof-of-principle measurements aimed at 
demonstrating the identification of a sufficiently large Cerenkov signal in homogeneous scintillating crystals to support dual-readout capability.  
}

\maketitle
\section{Introduction}
\label{intro}

The possibility of a future Higgs factory such as FCC-ee or CEPC motivates new and continuing efforts in the development of high resolution detector technologies.  A promising area of research is the development of high resolution calorimetry with state-of-the-art performance for both electromagnetic (EM) and hadronic signals.  The CalVision consortium seeks to advance homogeneous crystal and fiber calorimetry, and includes additional R\&D paths for new materials, sensors, readout, and full detector simulation and reconstruction to further enhance the physics capabilities of these new technologies.

A prominent research topic focuses on studies towards the design of a dual-readout (DR)~\cite{Akchurin:2005an} EM calorimeter using transversely segmented homogeneous crystals as both scintillators and Cerenkov radiators.  
Dual-readout calorimetry exploits the fact the relativistic particles produce Cerenkov (\v{C}) light, while the other forms of energy deposition only produce scintillation (S) light in scintillating optical calorimeters. By measuring both the \v{C} and S components, precise event-by-event calibrations can be used to extract a better energy measurement. 
The signal produced by a calorimeter shower initiated by an incident hadron is lower in general compared to that of an electron/photon of the same energy due to {\it invisible energy losses} from e.g. nuclear binding energies, neutrinos, and particles with small inelastic cross sections such as neutrons escaping the detector or depositing energy outside of the sampling window.  The amount of missing energy is correlated with the number of inelastic nuclear collisions. Calorimeter energy resolution can be improved by including measurable quantities that are correlated with the missing energy to make shower-by-shower corrections. An example is dual readout calorimetry, which uses scintillation light to measure the total ionizing energy deposit, and then timing and/or wavelength filters to collect Cerenkov light, which is produced only by relativistic particles in the shower (mainly electrons produced in $\pi^0$ showers).

\begin{figure}[th]
\centering
\includegraphics[width=5cm,clip]{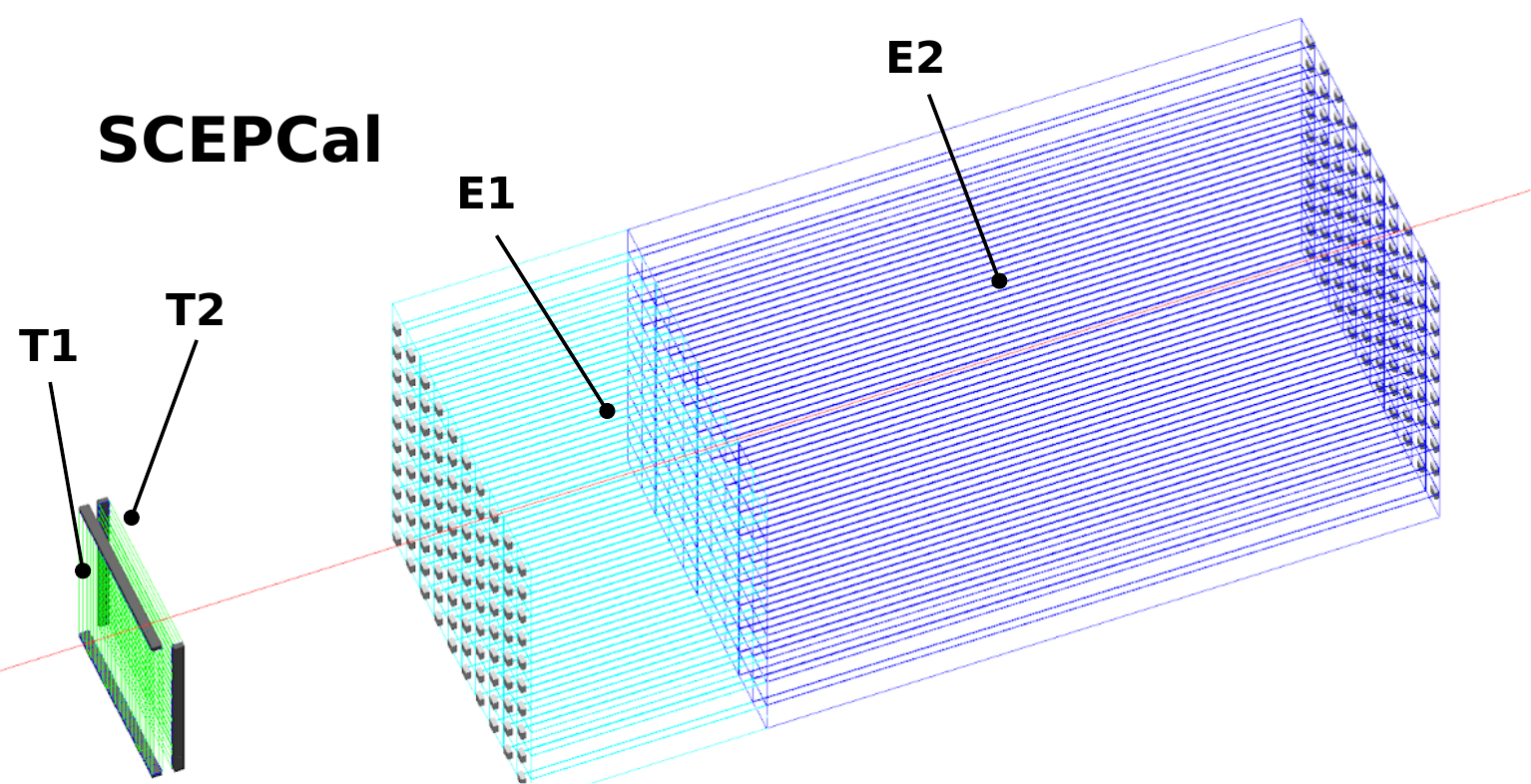}
\caption{Concept for a hybrid segmented calorimeter that exploits
scintillating crystals for detection of EM showers (figure adapted from Ref.~\cite{Lucchini:2020bac}).}
\label{fig:scepcal}       
\end{figure}

A detector concept, the Segmented Crystal Electromagnetic Precision Calorimeter (SCEPCal) is described in Ref.~\cite{Lucchini:2020bac} and illustrated in Fig.~\ref{fig:scepcal}.  This concept includes a hybrid segmented calorimeter using
scintillating crystals for detection of EM showers.  Optional layers providing precision timing for charged particles are shown as T1 and T2. The EM calorimeter is segmented into two layers: a front later, E1, with $\approx 6 \text{X}_0$ and finer granularity, and E2, with $\approx 16 \text{X}_0$ and DR capability.
In a full calorimeter system, the EM section could be followed by an ultrathin-bore solenoid and a DR hadron
calorimeter (HCAL) based on scintillating and quartz fibers, such as the one proposed by the IDEA collaboration~\cite{Antonello:2020tzq}. 

The challenges in implementing DR in a homogeneous scintillating detector element are the separation of the \v{C} signal and 
detection of a sufficient amount of these photons to apply the DR technique. Studies in Ref.~\cite{Lucchini:2020bac} indicate the need to collect more than 50 \v{C} photons/GeV 
to maintain the stochastic term of the calorimeter hadronic energy resolution below $28\%/\sqrt{E}$. 
Previous studies performed by the DREAM/RD52 programs~\cite{Akchurin:2012dw} validated the use of optical filters and waveform analysis to extract the \v{C} light component;  
however, the resulting \v{C} signals were found to be $2-3$ times lower than required.  
The detected \v{C} signal is affected by multiple factors, such as the short absorption length of (N)UV photons in scintillating crystals, the necessity to filter part of the spectrum before detection to remove a potentially overwhelming S signal, and the decreasing efficiency of many photodetectors at longer wavelengths.

We present results on the collection of Cerenkov and scintillation signals in PbWO$_4$ and \BGO crystal samples.  Compared to previous studies using a PMT readout, we employ modern SiPMs, which have significantly improved sensitivity to longer wavelength components of \v{C} signal which extend well beyond scintillation spectra are less affected internal absorption.  Results discussed include studies of modeling of S and \v{C} light collection, timing results for minimum ionizing particles (MIPs), and proof-of-principle measurements on the collection and identification of Cerenkov light signals for the realization of a homogeneous crystal EM layer with dual-readout capability.

\section{Experimental setup}

Two test beam campaigns were completed in the spring of 2023 using 120\,GeV protons at the Fermilab Test Beam Facility.  The primary goals of these initial studies were to evaluate the modeling of S and \v{C} light collection and to study the separation of the their signals in crystal scintillator samples.  A detailed study of the modeling of \v{C} light collection using a non-scintillating \PBF crystal was also performed and is reported in Ref.~\cite{Anderson:2024wqu}.

\begin{figure}[th]
\centering
\includegraphics[width=5cm,clip]{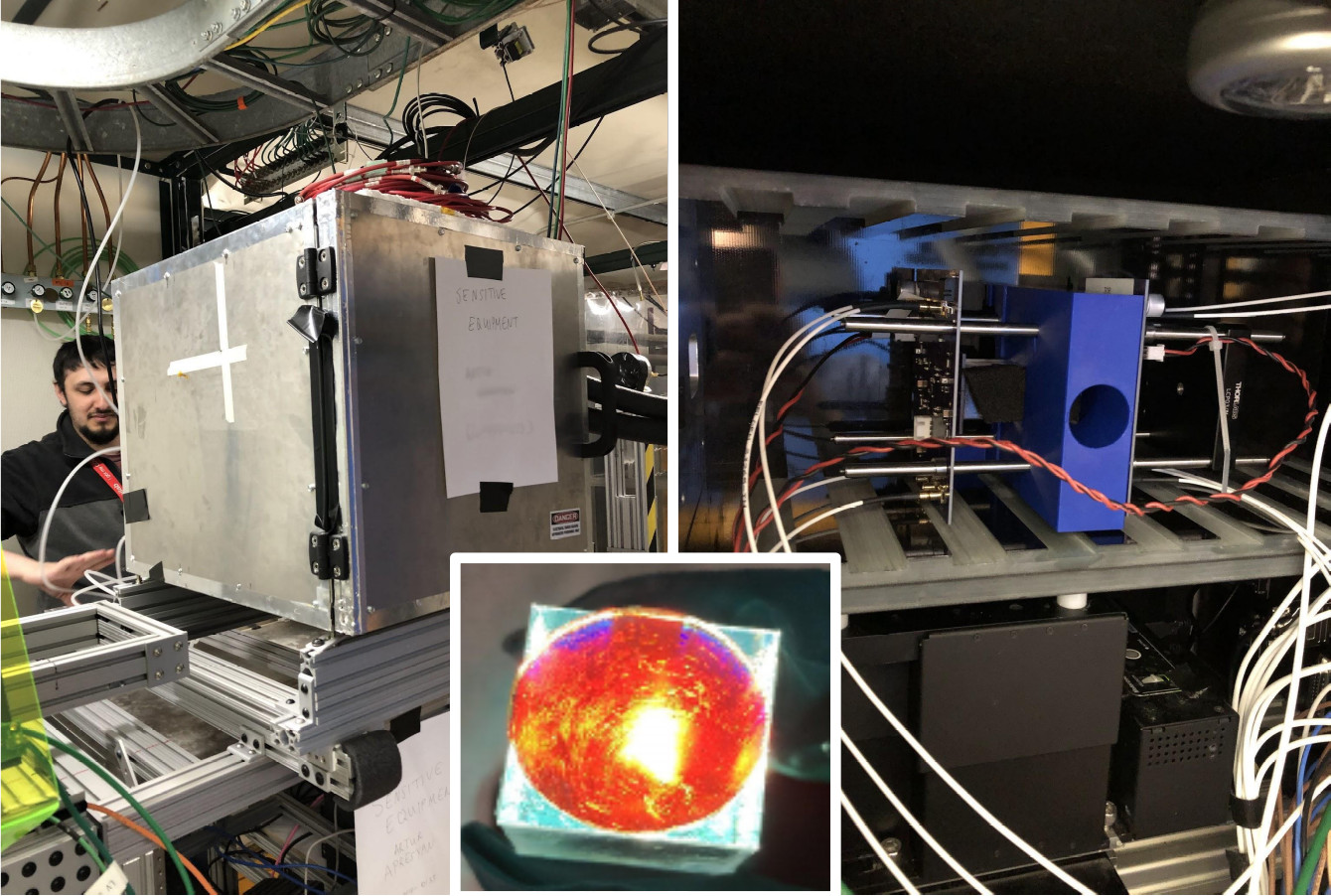}
\caption{Photo of the test set up.  The dark box and crystal mount with readout are shown with an inset photo of a 600\,nm long-pass optical filter mounted on a PbWO$_4$ crystal.  A silicon telescope upstream of the dark box provided beam tracking and an MCP positioned immediately downstream provided a precision time stamp with resolution $\approx 25$\,ps.}
\label{fig:TB}       
\end{figure}

Figure~\ref{fig:TB} shows the test beam configuration during the first campaign (see caption for additional details).  
Two crystal types were used in this test, \PWO and \BGO.  Both crystals were purchased from the Shanghai Institute of Ceramics, Chinese Academy of Sciences and have dimensions $25\text{mm}\times 25\text{mm}\times 60\text{mm}$. 
Crystals were mounted in a dark box with front and back faces instrumented using four 6\,mm$\,\times\,$6\,mm Hamamatsu 14160-6050HS SiPMs with a micro-cell pitch of $50 \mu$m. To improve light coupling between the crystals and SiPM windows, DOWSIL Q2-3067 optical grease was applied between the two surfaces.  The SiPM signals were amplified using the Infineon RF amplifier BGA616 with a 3 dB-bandwidth of 2.7 GHz and typical gain of 19.0 dB at 1.0 GHz. The AC-coupled SiPM signals and output of an MCP timing signal were digitized with a Lecroy Waverunner 8208HD oscilloscope at 10\,GS/s.

\section{Analysis of signals and light collection}
For each crystal, the collected light is expected to be dominated by scintillation light.  However, a Cerenkov light signal can in principle be detected by applying a combination of optical filters to suppress the scintillation signal and pulse shape analysis to identify the more prompt response expected for a Cerenkov signal.  The latter effect is a strong discriminant for scintillators with slower decay times, such as the BGO sample used in this test ($\tau > 300$\,ns).  The scintillation spectra peaks at 424\,nm (462\,nm) for the \PWO (\BGO)
samples and in each case the full-width, half-maximum (FWHM) of the spectrum is $\approx\,90$\,nm.
To suppress the scintillation component and enhance the relative proportion of \v{C} light detected, a  2.5 mm diameter, OD 2.0 longpass optical filter from Edmund Optics with cutoff centered at 600\,nm was used in \PWO runs, while a Schott UG11 notch filter with cutoffs centered at 380\,nm and 670\,nm was used in \BGO runs. Although the Cerenkov light spectrum is enhanced at UV wavelengths, the absorption length of the crystals can be quite short for these wavelengths and the effective cutoff for transmission may be close to or overlapping with the scintillation spectrum as is the case for \PWO. Therefore, detection of longer wavelength light is particularly important for measuring a \v{C} signal.

\subsection{Signal measurements}

\begin{figure}[t]
\centering
\includegraphics[width=.3\textwidth,clip]{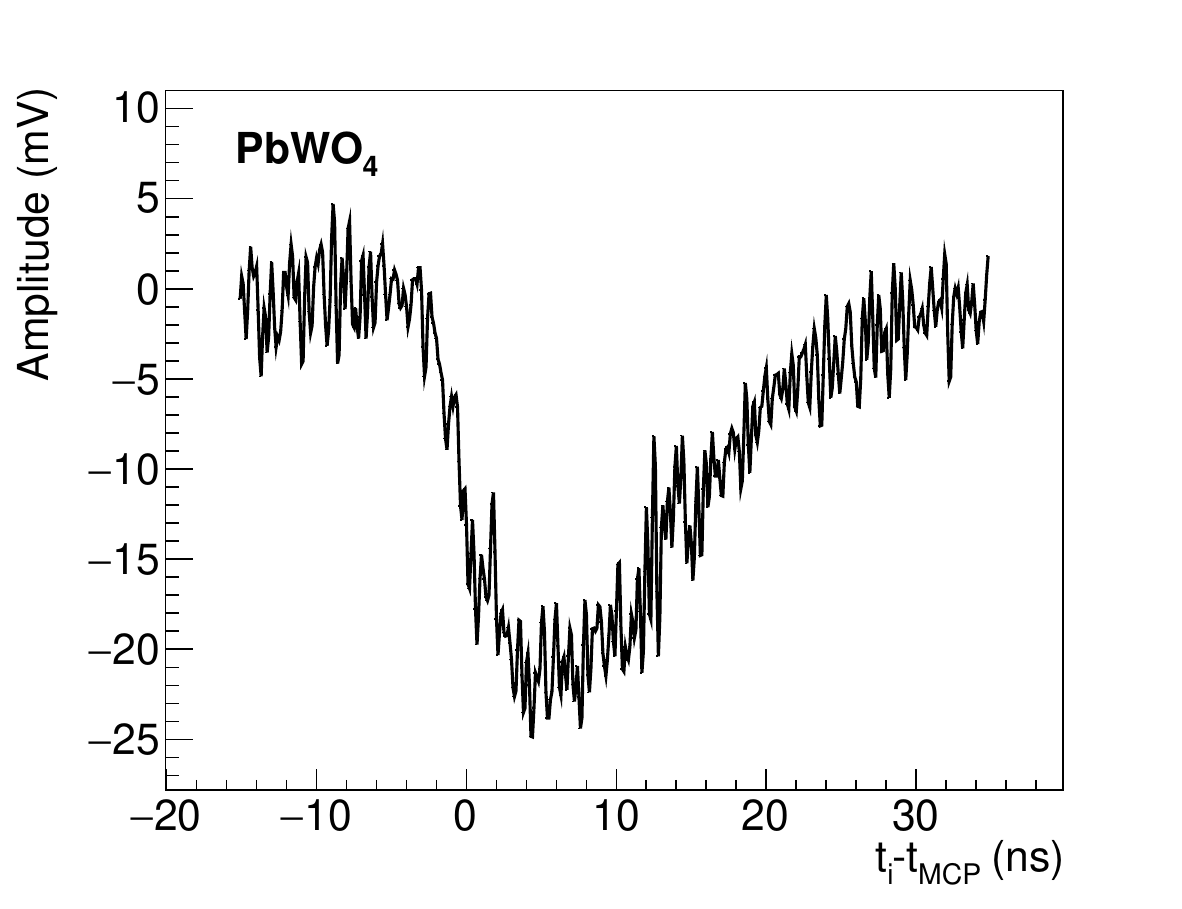}
\includegraphics[width=.3\textwidth,clip]{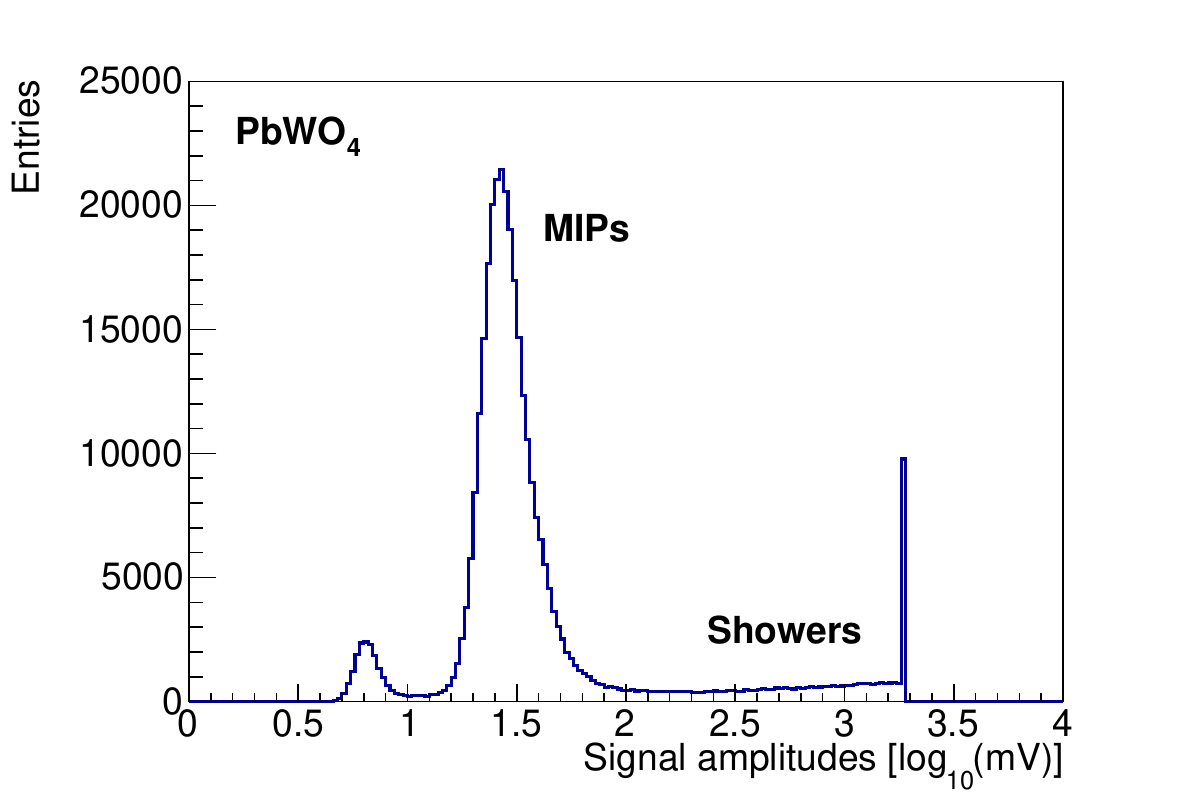}

\caption{Example of a digitized MIP signal (top) and the signal amplitude distribution on a log scale (bottom).}
\label{fig:signals}       
\end{figure}

Figure~\ref{fig:signals} (upper plot) shows a typical digitized waveform detected in a \PWO run with 120 GeV protons.  The protons interact primarily as minimum ionizing particles (MIPs), depositing only a few 10s of MeV. The samples are aligned relative to the MCP reference signal (\tMCP). 
The amplitude of the signal is evaluated as the maximum deviation of the waveform from the average amplitude of samples on the baseline ($t - \tMCP < -5$\,ns). The lower plot in Fig.~\ref{fig:signals} shows the observed distribution of amplitudes on a log$_{10}$ scale. The first smaller peak at 0.7 corresponds to events where protons miss the crystal and reflects the noise level. The main peak at 1.4 is due to MIP tracks that are well contained within the transverse dimensions of the crystal.  Above the MIP peak part of the distribution for showering particles is evident until reaching saturation for this readout configuration.

\subsection{Timing resolution studies}

\begin{figure}[ht]
\centering
\includegraphics[width=0.22\textwidth,clip]{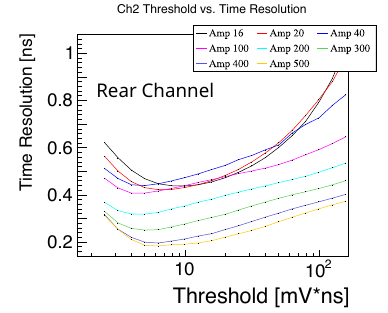}
\includegraphics[width=0.23\textwidth,clip]{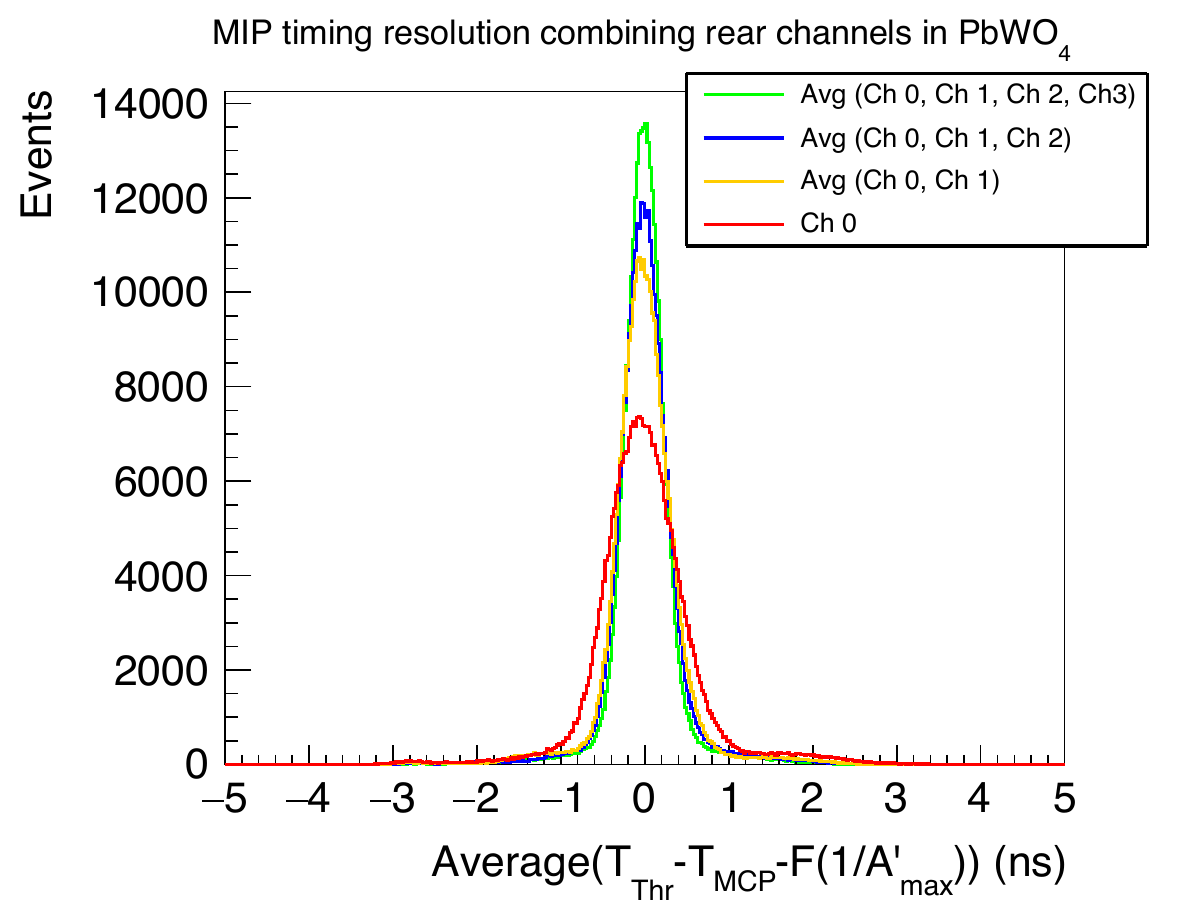}
\caption{Time resolution versus threshold for single rear channel in PbWO$_4$ (left). Average time resolution when combining rear channels in the PbWO$_4$ for MIPs (right).}
\label{fig:PWOtime}       
\end{figure}

Figure~\ref{fig:PWOtime} presents selected measurements of the timing performance in \PWO. The left plot of considers the timing resolution for a single rear SiPM channel as a function of signal amplitude.   Events are selected in narrow bins of pulse amplitude to avoid significant time walk effects and signal times are defined by integrating the pulses until a certain threshold requirement is satisfied.  The timing resolution is plotted for different signal amplitudes and as a function of the threshold requirement.  The right plot shows the timing resolution for MIP signals in the PbWO$_4$ crystal after successively combining the signals from the 4 SiPMs on the rear face of the crystal. The optimal threshold for the average MIP amplitude was used to define the timing. Time walk corrections are applied to compensate for the finite rise time of the signals, and to allow for their combination.  A single channel timing resolution for the MIP signals is found to be $\approx 420$\,ps, while the combined timing resolution is $\approx 225$\,ps. Since the time resolution improves with amplitude, the timing resolution can be expected to improve with a larger crystal where more of the shower can be contained.

\subsection{Modeling light collection}

\begin{figure}[th]
\centering
\includegraphics[width=0.48\textwidth,clip]{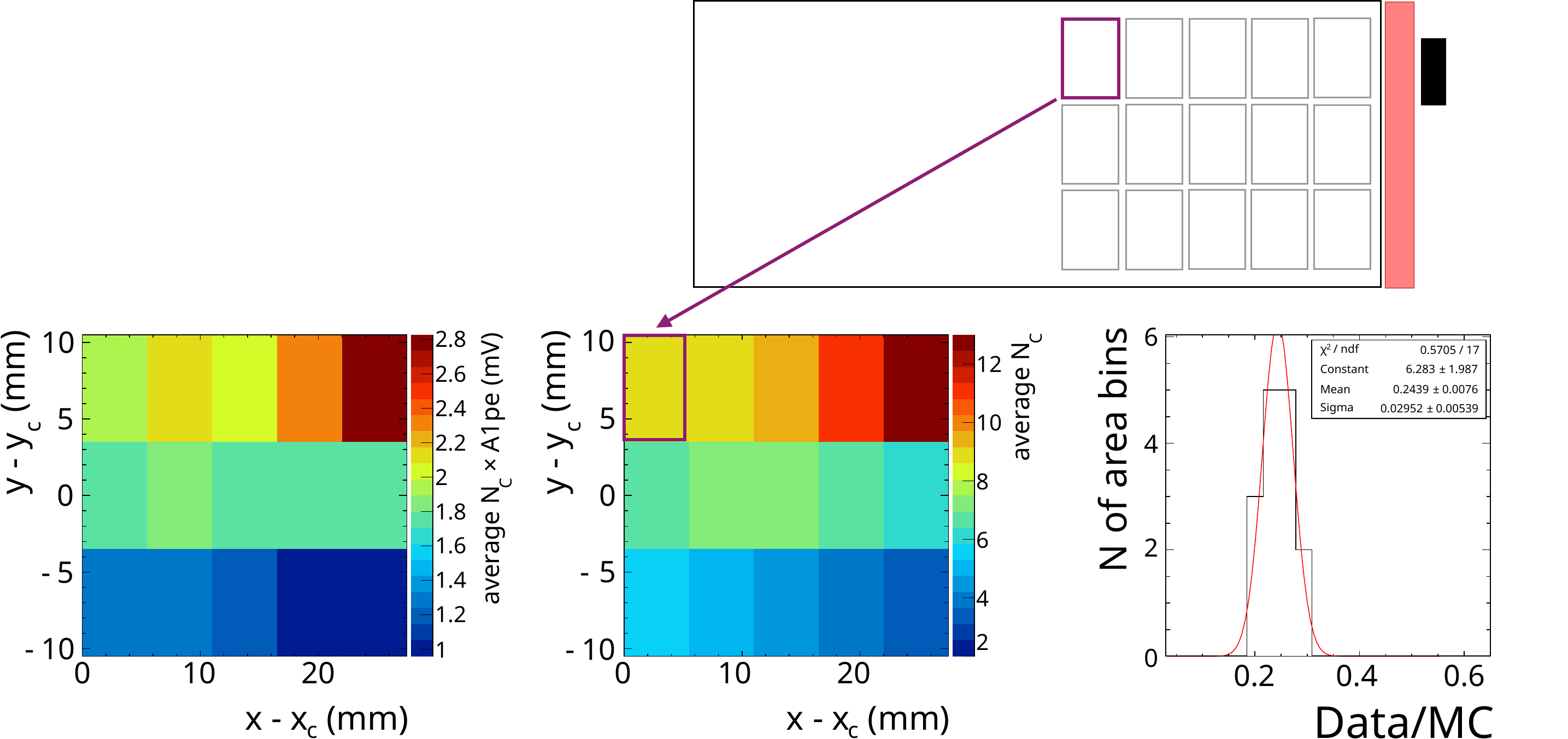}
\caption{Test of Geant4 modeling of Cerenkov light collection in the BGO crystal.  See text for details. }
\label{fig:BGOmodel}       
\end{figure}

One of the goals of the test beam program is to study the modeling of the production and collection of optical photons by Geant4~\cite{GEANT4:2002zbu} to validate its use for future optimizations or our DR detector.
Figure~\ref{fig:BGOmodel} shows a comparison of beam data to a simulation of our BGO crystal in Geant4 with the optical photon processes fully modeled.  In the simulation study, the beam was incident transverse to the long axis of the crystal and the signal was read out by two SiPMs on the top-right rear face of the crystal shown as a black box in the drawing.  The measurement included the UG11 filter to preferentially select Cerenkov light for this study.  The side of the crystal facing the beam is divided into rectangular regions.  The color maps show the measured (left) average signal amplitude detected in the SiPMs for protons incident in each region and (right) the number of photons propagated into the SiPMs in a Geant4 model of the experiment.  The histogram shows the bin-by-bin ratio of data divided by simulation.  The mean value of the histograms represent a calibration factor including effects of SiPM PDE and losses in light due to coupling at optical interfaces that are not well modeled.  
The relative spread of $\approx 10\%$ in modeling the distribution of collected light over the regions, which is also observed repeating the analysis for S light, gives confidence that the simulation is reproducing the overall details of the light collection relatively well.  This is also the case in our recent study presented in Ref.~\cite{Anderson:2024wqu}.

\subsection{Cerenkov light separation and yield}

\begin{figure}[th]
\centering
\includegraphics[width=0.23\textwidth,clip]{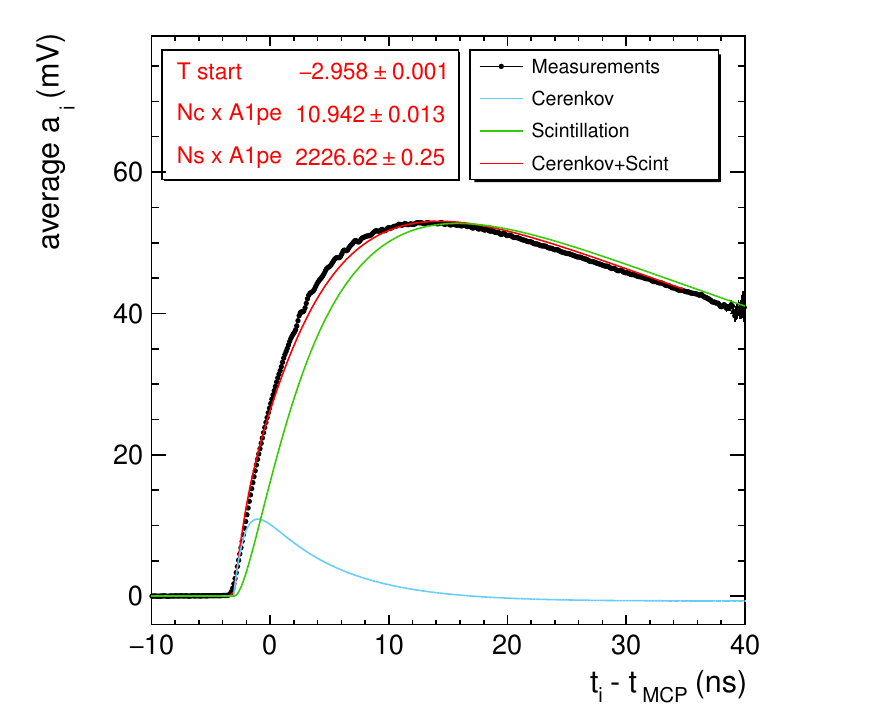}
\includegraphics[width=0.23\textwidth,clip]{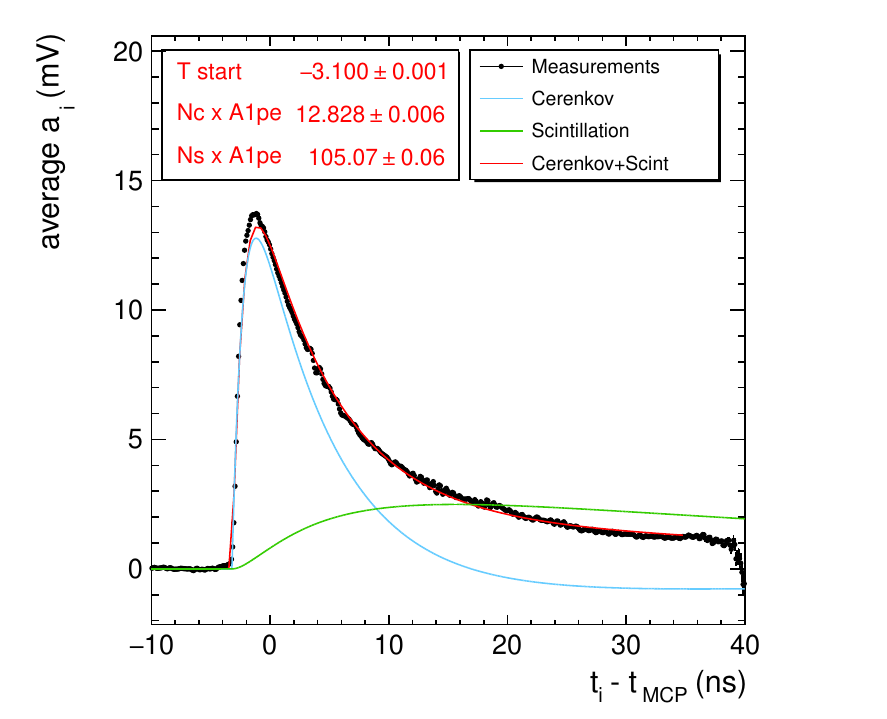}
\caption{Average waveforms measured for MIPs in the BGO data without (left) and with (right) the UG11 filter placed in front of the SiPMs. See text for details.}
\label{fig:SandCMIP}       
\end{figure}

The performance of DR calorimetry critically depends on the identification of S and \v{C} light signals associated with particle showers.  A comprehensive review of the DR concept can be found in Ref.~\cite{Lee:2017xss}.  Average waveforms measured for MIPs in the BGO data are shown in Fig.~\ref{fig:SandCMIP} without (left) and with (right) the UG11 filter placed in front of the SiPMs.  A template fit is applied to the pulses to determine the relative contribution of \v{C} and S signals.  The templates are generated as follows. For the \v{C} signal template the impulse response of the SiPM+amplifier, generated using an O(100)\,ps laser pulse, is convoluted with the expected distribution of the arrival time of photons at the SiPM surface from Cerenkov radiation.  This distribution is calculated using a detailed optical simulation of the optical photons in the crystal.
The S signal template begins with the impulse response of the SiPM+amplifier and is then convoluted with the measured decay time of the \BGO scintillation.  The decay time is much longer than the light propagation time in the crystal, therefore it is not necessary to perform a detailed ray tracing procedure.  
Because of the high S light yield in \BGO this waveform agrees very well with the one measured with no filter applied.   The model and wave forms in data show agreement to within a few percent.

\begin{figure}[th]
\centering
\includegraphics[width=5cm,clip]{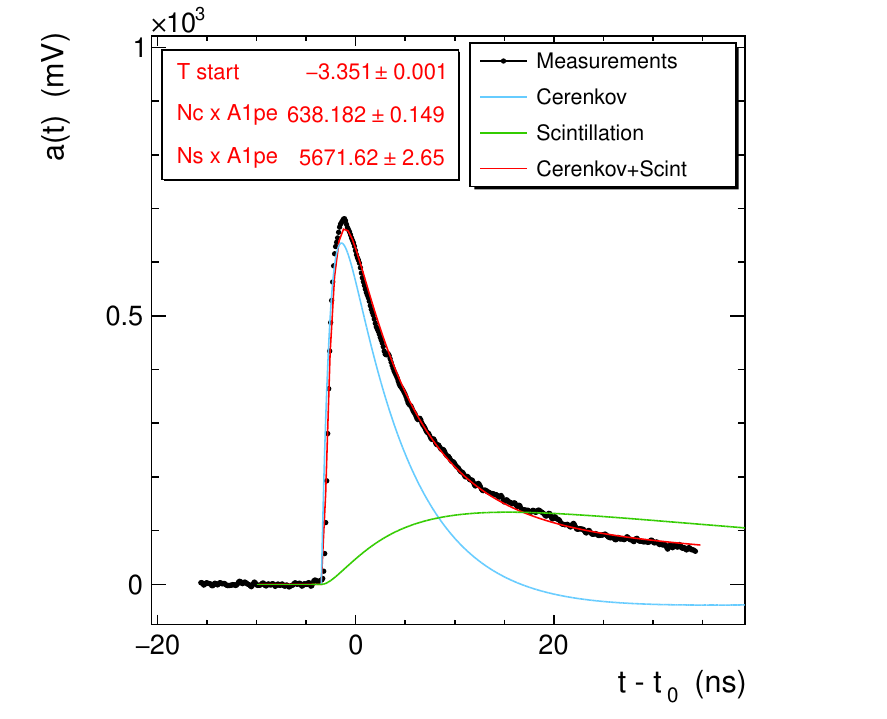}
\caption{Fit for S and \v{C} signal components applied to a proton shower event.}
\label{fig:BGOshower}       
\end{figure}

The amplitudes of the components are interpreted as the count of detected photons multiplied by the response of the  SiPM+amplifier to a single detected photon.  The single photon response (SPR) is determined using a highly attenuated O(100)\,ps laser pulse and found to be $\approx 0.7$\,mV for the bias setting used in these data.  The measured signals are normalized to the MIP energy deposition spectrum calculated in Geant4.  Compared to showers the MIP response was found to be more strongly affected by the relative position of tracks to the SiPMs when traversing the long axis of the crystal.  To extract the \v{C} signal from MIPs we instead used the transverse measurements shown in Fig.~\ref{fig:BGOmodel}.  From a mean energy deposition of 25.4 MeV, calculated in the simulation, we measure a Cerenkov light output (n\v{C}) of 5.2 photons or 203 n\v{C}/GeV.  A correction for the angle of incidence using an optical simulation in Geant4 gives $\approx 300$ n\v{C} per GeV of deposited energy for protons traversing the long axis. We estimate an uncertainty of [+10,-20]\% on the SPR calibration in these runs.  Figure.\ref{fig:BGOshower} shows the fitting method applied to a large larger amplitude showering event.  This is the first proof of principle measurement for the viability of applying DR in a homogeneous crystal EM calorimeter. 

\section{Summary}

A selection of results from the first CalVision test beam are presented.  An analysis of MIPs in a 6\,cm \PWO crystal shows a timing resolution approaching $\Delta t=200$\,ps.  Simulations  of the production and collection of Cerenkov light were validated in comparisons with test beam data.  Our measurements of the \v{C} signal with a SiPM-based readout indicate the identification of significantly more than 50 \v{C} photons/GeV 
needed to maintain an optimal stochastic term for hadrons in a dual readout calorimeter, providing the first proof of principle for the viability of applying DR in a homogeneous crystal EM calorimeter.  Future studies will focus on additional measurements with improved calibration for photon counting and reduced dependence on simulation.

\section{Acknowledgments}
\label{ackno}
This work was supported via U.S. Department of Energy Grant DE-SC0022045. The test beam was done with support from Artur Apresyan and the facilities of the Compact Muon Solenoid Experiment Endcap Timing Layer. Some of the work presented was prepared  using the resources of the Fermi National Accelerator Laboratory (Fermilab), a U.S. Department of Energy, Office of Science, Office of High Energy Physics HEP User Facility. Fermilab is managed by Fermi Research Alliance, LLC (FRA), acting under Contract No. DE-AC02-07CH11359. 

%
%
%

\end{document}